\documentclass[runningheads]{llncs}
\usepackage{graphicx}
\usepackage{xcolor}
\usepackage{algorithm}
\usepackage{algpseudocode}
\usepackage{fancyhdr}
\fancypagestyle{mystyle}{
	\fancyhf{}
	\fancyhead[RO,LE]{\thepage}

}

\begin{document}
\thispagestyle{mystyle}

\title{Blockchain in Healthcare: Implementing Hyperledger Fabric for Electronic Health Records at Frere Provincial Hospital} 
\titlerunning{Blockchain in Healthcare: Implementing Hyperledger Fabric for...}
%
\author{Olukayode Ayodele Oki\inst{1}\orcidID{0000-0002-6887-9782} \and \\
Abayomi O. Agbeyangi\inst{1}\orcidID{0000-0002-2504-2042} \and \\
Aphelele Mgidi\inst{1}\orcidID{0000-0002-1085-6465}}

\authorrunning{O.A. Oki and A.O. Agbeyangi}
%
\institute{Walter Sisulu University, Buffalo City South Africa, 
\email{\{ooki,aagbeyangi\}@wsu.ac.za}}
\maketitle              
\begin{abstract}
As healthcare systems worldwide continue to grapple with the challenges of interoperability, data security, and accessibility, integrating emerging technologies becomes imperative. This paper investigates the implementation of blockchain technology, specifically Hyperledger Fabric, for Electronic Health Records (EHR) management at Frere Hospital in the Eastern Cape province of South Africa. The paper examines the benefits and challenges of integrating blockchain into healthcare information systems. Hyperledger Fabric's modular architecture is harnessed to create a secure, transparent, and decentralized platform for storing, managing, and sharing EHRs among stakeholders. The study used a mixed-methods approach, integrating case studies and data collection methods through observation and informal questions, with the specific goal of understanding current record management methods and challenges. This method offers practical insights and validates the approach. The result demonstrates the role of blockchain in transforming healthcare, framed within a rigorous exploration and analysis. The findings of this study have broader implications for healthcare institutions seeking advanced solutions to address the persistent challenges in electronic health record management. Ultimately, the research underscores the transformative potential of blockchain technology in healthcare settings, fostering trust, security, and efficiency in the management of sensitive patient data.

\end{abstract}
\section{Introduction}
Blockchain technology has expanded beyond its original connection with financial technology and is now making its way into different areas, such as healthcare \cite{Khezr2019,Massaro2023}. Blockchain is recognised for its ability to disrupt banking institutions and its close association with cryptocurrencies. It is now being seen as a promising way to improve data security, transparency, and interoperability in healthcare environments \cite{Yaqoob2021}. Initially, cryptocurrencies like Bitcoin relied on blockchain as their foundational technology. It functions as a decentralised and unchangeable system of record-keeping. By utilising cryptographic techniques, it ensures transparency, security, and tamper-resistance when storing transactions or data records across a dispersed network of nodes. According to Tseng and Shang \cite{Tseng2021}, it removes the necessity for intermediaries seen in conventional centralised databases, enabling direct transactions between peers while preserving the integrity of data.

Managing electronic health records (EHR) in the traditional healthcare system encounters notable hurdles such as data security risks, interoperability problems, and administrative inefficiencies. The problems impact patient privacy, impede smooth information sharing between healthcare professionals, and lead to inefficiencies in healthcare service delivery. Despite attempts to tackle these issues with technical progress, current solutions have not fully achieved widespread and lasting enhancements in EHR management.
Hyperledger Fabric \cite{Androulaki2018}, an open-source blockchain platform designed for enterprise use, shows great potential for transforming multiple industries, such as healthcare. Its distinctive attributes make it especially suitable for tackling the issues found in conventional healthcare information management systems \cite{Sutradhar2024}. By providing unique benefits compared to other blockchain frameworks and conventional data management methods, it is suitable in healthcare settings. For example, Hyperledger Fabric's permissioned blockchain network allows for greater control over who can access and contribute to the stored data, ensuring privacy and security. By harnessing the power of this innovative technology, healthcare facilities can enhance the quality of care they provide while reducing costs and increasing patient trust.

The motivation for the study is due to the benefits of this technological innovation as well as its security, technical aspects, and operational feasibility at healthcare facilities. The current increase in specialised healthcare services and patient movement emphasises the crucial need to provide healthcare facilities with thorough patient medical histories. Sharing clinical data securely across several healthcare facilities is a complex task that demands a careful equilibrium between confidentiality, data integrity, and patient privacy.  

The main contribution of the paper is to explore the transformative potential of implementing a blockchain-based Electronic Health Record (EHR) system. aiming to revolutionise healthcare data management and improve patient care outcomes. Other specific contributions of the paper include:
\begin{itemize}
	\item examining the viability, advantages, and obstacles of implementing a blockchain-based EHR system
	\item the implementation of the blockchain-based EHR system.
\end{itemize}

The Frere Hospital, where the study is conducted as a case study, is a provincial and public hospital, and it plays a vital role as the primary medical facility catering to the healthcare needs of East London and the broader Eastern Cape region in South Africa. It focuses on providing essential medical services to underprivileged communities, ensuring access to quality healthcare for all. By introducing a blockchain-based solution, the study seeks to demonstrate the potential benefits of blockchain technology in improving data security, integrity, and accessibility in healthcare settings. While the system is initially an ongoing research project, its ultimate goal is to showcase its feasibility and effectiveness in real-world healthcare environments and potentially pave the way for adoption.

The remaining part of the paper is structured as follows: Section 2 discusses the literature review; Section 3 explains the methods; the discussion on the implementation is in Section 4; and Section 5 concludes the paper.

\section{Literature review}
\subsection{Background}
Blockchain has emerged as a very significant and rapidly evolving subject on a global scale, especially in the context of financial technologies, since the start of the 21st century. This technology, in conjunction with distributed database technologies, plays a crucial role in facilitating advancements in distributed transaction and ledger systems, hence creating fresh possibilities for digital platforms and services \cite{Dutta2020,Gad2022}. According to Bhutta et al. \cite{Bhutta2021}, the blockchain environment consists of a decentralised system that uses cryptography to record and store an unchangeable, reliable, and sequential log of transactions amongst interconnected participants. Similar to a distributed ledger, the parties involved maintain, update, and validate this system. This approach eliminates the necessity of a central authority to authenticate transactions, as all members of the network collectively verify and safeguard the information \cite{Aggarwal2019}. The extensive implementation of blockchain technology across multiple sectors, such as finance\cite{Zhang2020}, supply chain management\cite{Dutta2020}, healthcare\cite{Massaro2023}, and voting systems\cite{Hjalmarsson2018}, is mostly due to its high level of transparency and security. The potential uses of blockchain technology are extensive, with certain experts forecasting that it has the ability to fundamentally transform the methods by which we engage in commerce and transfer assets in the coming years.

The healthcare management sector is constantly changing as new technical advancements provide fresh ways to tackle long-standing challenges. Blockchain technology's implementation in healthcare represents a fundamental shift in the management of electronic health records (EHRs) and patient data \cite{Yaqoob2021}. Although traditional EHR systems have been effective in converting medical records into digital format and improving administrative efficiency, they frequently encounter challenges with the accuracy and security of data, as well as the capacity to seamlessly exchange information with other systems. Menachemi and Collum (2011) opined that EHR systems face some challenges, including high upfront costs, ongoing maintenance expenses, workflow disruptions leading to temporary productivity losses, and potential privacy concerns among patients. Specifically, blockchain provides a secure and verifiable record of patient data exchanges, which helps address these difficulties. It enhances data security and confidentiality by decentralising data storage and employing cryptographic techniques to prevent unauthorised access, data breaches, and manipulation.

Furthermore, the distributed architecture of blockchain allows for effortless compatibility between different healthcare systems and participants, facilitating the safe transfer of patient information between different organizations. Interoperability is especially vital in healthcare settings, as patient care frequently requires cooperation across many clinicians and institutions. The lack of interoperability, as stated by Iroju et al. \cite{Iroju2013}, poses significant challenges in healthcare systems, where health practitioners may face difficulties in obtaining complete patient information, leading to repeated tests and procedures. Healthcare organisations may optimise data sharing processes, boost care coordination, and improve patient outcomes by utilising blockchain technology. Also, blockchain technology has the potential to empower patients by giving them more authority over their health data, enabling consent management, and promoting transparency in healthcare transactions \cite{Esmaeilzadeh2019}.

Although blockchain has great promise, its implementation in the healthcare industry is not without obstacles. According to Sharma and Joshi \cite{Sharma2021}, low awareness of legal issues and inadequate support from high-level management are the most influential driving barriers, based on their study on blockchain adoption in the Indian healthcare industry. However, the investigation of blockchain technology in healthcare gives a promising chance to tackle the intricacies of EHR management, boost data security and interoperability, and eventually improve the provision of healthcare services to patients.

\subsection{Related Work}
In healthcare, studies have been conducted on the impact and feasibility of blockchain to enhance healthcare service delivery \cite{Sutradhar2024,Pandey2020,Zarour2020,Radanović2018,Antwi2021,Stamatellis2020,Esmaeilzadeh2019,Sharma2021}. Several of these studies have shown promising results, particularly in enhancing patient data security and improving interoperability among different healthcare providers. It was stated by Pandey and Litoriya \cite{Pandey2020} that blockchain provides a secure and transparent system for integrated healthcare services, ensuring corruption-free and efficient implementation of nationwide health insurance programs. Furthermore, research has also highlighted the potential of blockchain technology in reducing medical errors \cite{Radanović2018,Zarour2020} and improving overall patient care quality. Overall, the existing literature emphasises the importance of analysing and designing blockchain-based EHR systems to revolutionise healthcare delivery. 
An important feature of blockchain technology in healthcare, specifically using Hyperledger Fabric, is its capacity to guarantee the integrity and confidentiality of data \cite{Antwi2021}. This, in turn, improves patient privacy and security. In the study by Sutradhar et al. \cite{Sutradhar2024}, the paper presents a specialised framework for managing identity and access in the healthcare industry using blockchain technology. By utilising Hyperledger Fabric and OAuth 2.0, it was shown that the framework guarantees improved security and scalability while also providing transparency, immutability, and fraud prevention. 

Another interesting approach was by Antwi et al. \cite{Antwi2021}; the study examines the practicality of using private blockchain technologies, notably Hyperledger Fabric, to address different requirements and scenarios in the healthcare industry. Empirical assessments in the study reveal the significant advantages of the approach, such as heightened security, adherence to regulations, interoperability, adaptability, and scalability, hence demonstrating their potential to efficiently tackle crucial issues in healthcare data management. The study by Stamatellis et al. \cite{Stamatellis2020} also looked at the changes that have happened in healthcare from using paper-based systems to electronic health records (EHRs) and the problems that have come with them, mainly looking at cybersecurity risks like malware and ransomware. It specifically focuses on notable events like the WannaCry ransomware cryptoworm and the Medjack cyberattack. The research highlights the susceptibility of conventional EHR databases to typical methods of attack and emphasises the necessity for a scalable, unchangeable, transparent, and secure solution to tackle these difficulties. The results of the experimental evaluation provide evidence of the advantages of the method in improving security, compliance, compatibility, flexibility, and scalability. This offers a potential alternative for managing decentralised medical data, as also implied in this research.

Similarly, an effort by Al-Sumaidaee et al. \cite{Al-Sumaidaee2023} addresses the difficulties in the healthcare sector caused by inadequate interoperability and fragmented communication systems, which have negative effects on patients, resources, and costs. The paper suggests implementing blockchain technology, specifically Hyperledger Fabric, to enhance the exchange of information among various healthcare organisations by introducing a trust element. It emphasises the substantial influence of factors such as workforce size and transactions per second (TPS) on network latency and overall throughput.
The studies emphasise the potential of blockchain technology, specifically Hyperledger Fabric, in strengthening healthcare service delivery, improving patient data security, minimising medical errors, and transforming healthcare practices. Although recent studies show encouraging outcomes, there are still unresolved concerns that need to be tackled. Some of the challenges include scalability, regulatory compliance, interoperability, and the requirement for additional empirical validation in real-world healthcare environments. This study conducted empirical assessments and real-world implementations to provide practical insights and validation for the solutions, considering these issues. It is aimed at enhancing patient outcomes and healthcare service delivery by raising awareness and acceptance of blockchain technology in the healthcare sector.

\section{Methods} 
The study uses a mixed-methods approach, starting with requirement gathering, to investigate the adoption of blockchain technology, specifically the Hyperledger Fabric, for the management of electronic health records (EHR) at Frere Provincial Hospital. The implementation was divide into separate components, each of which fulfils a specific role in the system's functionality. The main constituents consist of the Hyperledger Fabric network and distributed ledger, smart contracts, the application Software Development Kit (SDK), and the application frontend.

\subsection{Requirement Gathering}
The research collected data as part of the requirement gathering for the system implementation. The goal was to gather patient insights while also observing the operational dynamics at Frere Provincial Hospital. A structured questionnaire was designed as a form of standardised format for data collection. To encourage open and honest feedback, the respondents were assured of the anonymity of their responses. It was distributed to a sample group of patients, encompassing a diverse demographic. This was to investigate concerns related to the current non-digital storage of medical records and assess the perceived impact of administrative processes on the overall service delivery experience. 

For the on-site operational observations, physical observation was conducted in the hospital environment, paying attention to document handling, data retrieval, and administrative workflows. This was followed by an informal interview with the hospital staff to gather qualitative insights into the challenges faced in data management. Prior informed consent was obtained from participants in adherence to ethical considerations without collecting personally identifiable information.

\begin{figure}[]
	\centering
	\includegraphics[width=3in]{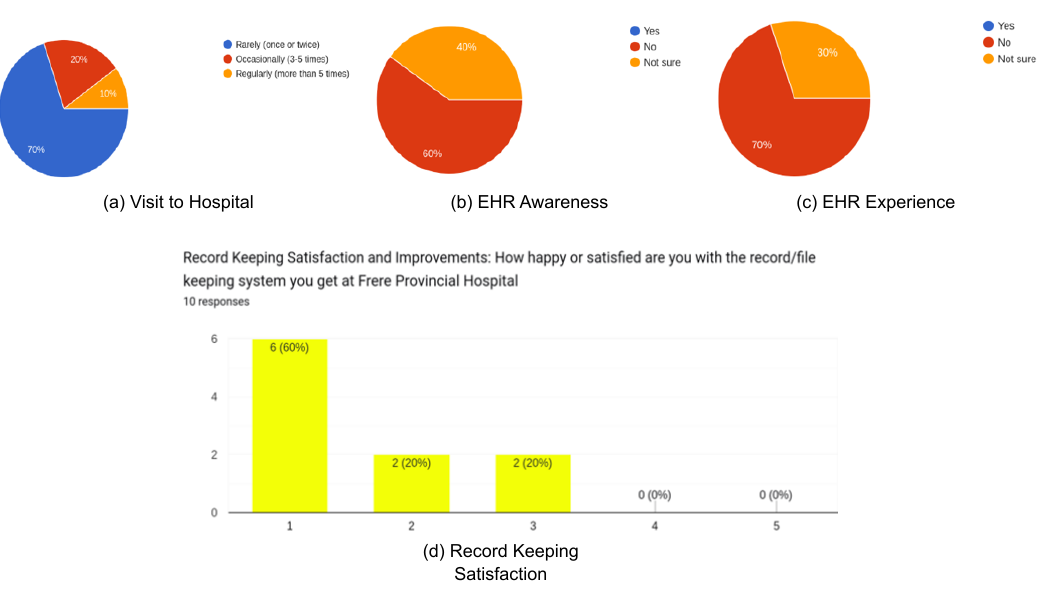}
	\caption{Insights on Electronic Health Record System: (a) Frequency of hospital visits among patients (b) Awareness of Electronic Health Records (EHR) (c) Experience with using EHR systems (d) Current record keeping satisfaction.} \label{fig1}
\end{figure}

Figure \ref{fig1} provides a visual representation of the requirement-gathering analysis, based on responses from a random sample of 10 patients. Subfigure \ref{fig1}(a) shows the frequency of patient visits to the hospital, with 70\% of respondents visiting regularly (more than five times), indicating a high reliance on the hospital's services. Subfigure \ref{fig1}(b) illustrates the awareness of electronic health records (EHR) among patients, revealing that 60\% of respondents are not aware of EHRs, suggesting a significant gap in patient knowledge about digital health record systems. Subfigure \ref{fig1}(c) depicts the experience of patients with EHRs; 70\% of respondents have no prior experience with EHRs, highlighting the lack of exposure and potential challenges in transitioning from a non-digital to a digital record-keeping system. Finally, subfigure \ref{fig1}(d) reveals that 60\% of respondents are dissatisfied with the current manual record-keeping system, pointing to a critical need for improvement in the hospital's data management processes. This data underscores the necessity for implementing a blockchain-based EHR system to enhance efficiency, security, and patient satisfaction.

\subsection{The Hyperledger Fabric Components}
The Hyperledger Fabric's components are crucial and provide the technical foundation for understanding the application of the proposed blockchain technology implementation for electronic health records (EHR).

\begin{figure}[ht]
	\centering
	\includegraphics[width=3in]{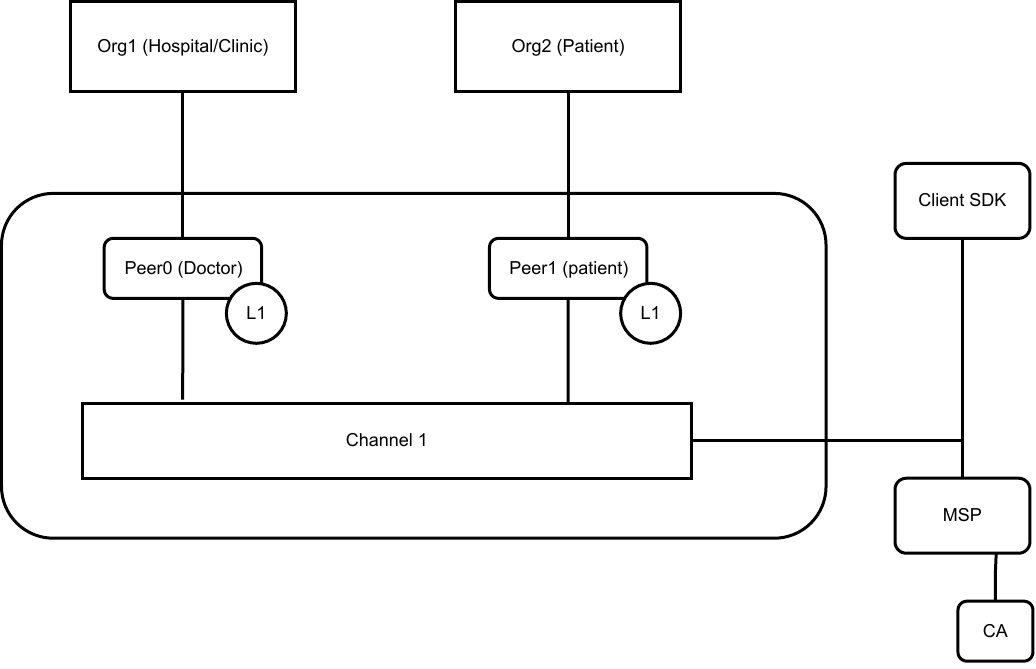}
	\caption{Overview of the Hyperledger Fabric Architecture} \label{fig2}
\end{figure}

The modular architecture of the Hyperledger Fabric, as described in Figure \ref{fig2}, is specifically designed to offer a flexible and scalable framework for blockchain solutions, forming the backbones of the entire application, managing the ledger, and facilitating interactions between various participants. The smart contracts are crafted using JavaScript, a versatile and widely used programming language. These contracts define the transactional logic and rules governing the interactions between different entities within the Hyperledger Fabric network. The front end of the application used Angular, a popular web application framework. Angular\footnote{https://angular.io/} provides a robust and dynamic user interface, enabling a user-friendly experience. It serves as the interface through which users (patients, doctors, and administrators) interact with the system. The test network adaptation entails transforming organizations within the network to represent distinct hospitals. This modification entails adjusting Docker files, configuration files, and corresponding certificates. Specifically, alterations are made to organization names in the \textit{``configtx.yaml"} file, and adjustments are applied to the certificate authorities in the docker-compose file. Afterwards, the files that reference these configurations are updated. Once this custom network is set up and a designated channel is established, as shown in the architecture, the necessary certificates are made for the organisation and peers that are involved.
 
Also, as illustrated in Figure \ref{fig2}, the Hyperledger Fabric architecture comprises two main organisations: Org1 (hospitals or clinics) and Org2 (patients). Each organisation has peers—Peer0 for doctors and Peer1 for patients—connected via Channel 1, which ensures secure, transparent, and verifiable data exchanges. Both peers maintain a ledger (L1) that records all transactions. The Client SDK facilitates application interactions with the blockchain network, while the Membership Service Provider (MSP) handles identity management, ensuring authorised access. The Certificate Authority (CA) issues digital certificates for identity validation, enhancing the system's security. This architecture supports secure, efficient data sharing, giving patients control over their health records, and enabling healthcare providers to update patient information as needed.

\subsection{Smart contract implementation and role-based access control}
In line with the application's modularity, all components are pluggable. The backend code and smart contracts  utilise \textit{ExpressJS} as a server to deliver REST APIs. Communication between the frontend and backend is facilitated through REST calls, incorporating JSON web tokens for authentication. he fabric framework supports two databases - \textit{LevelDB} and \textit{CouchDB}. The \textit{CouchDB} database was chosen due to its enhanced flexibility, especially in handling images and supporting indexes, and features vital for the healthcare context. Given that all patient data resides within the \textit{CouchDB}, obviating the need for a separate Electronic Health Record (EHR) store, \textit{CouchDB} aligns perfectly with the system's requirements. In this architecture, \textit{CouchDB} is leveraged to store the world state. This design choice eliminates the need to query the entire transaction log for each transaction request, streamlining system efficiency. By providing quicker access to relevant information without traversing the entire transaction history, \textit{CouchDB} optimises the overall performance of the system.

In the application, all the executable business logic is encapsulated within smart contracts. This implies that any operations involving the creation, retrieval, update, or deletion of records in the distributed ledger are orchestrated through smart contracts. Each functionality required by the system to interact with the Hyperledger Fabric (HLF) network is encapsulated within a separate function for clarity and ease of maintenance.

\begin{figure}[ht]
	\centering
	\includegraphics[width=3in]{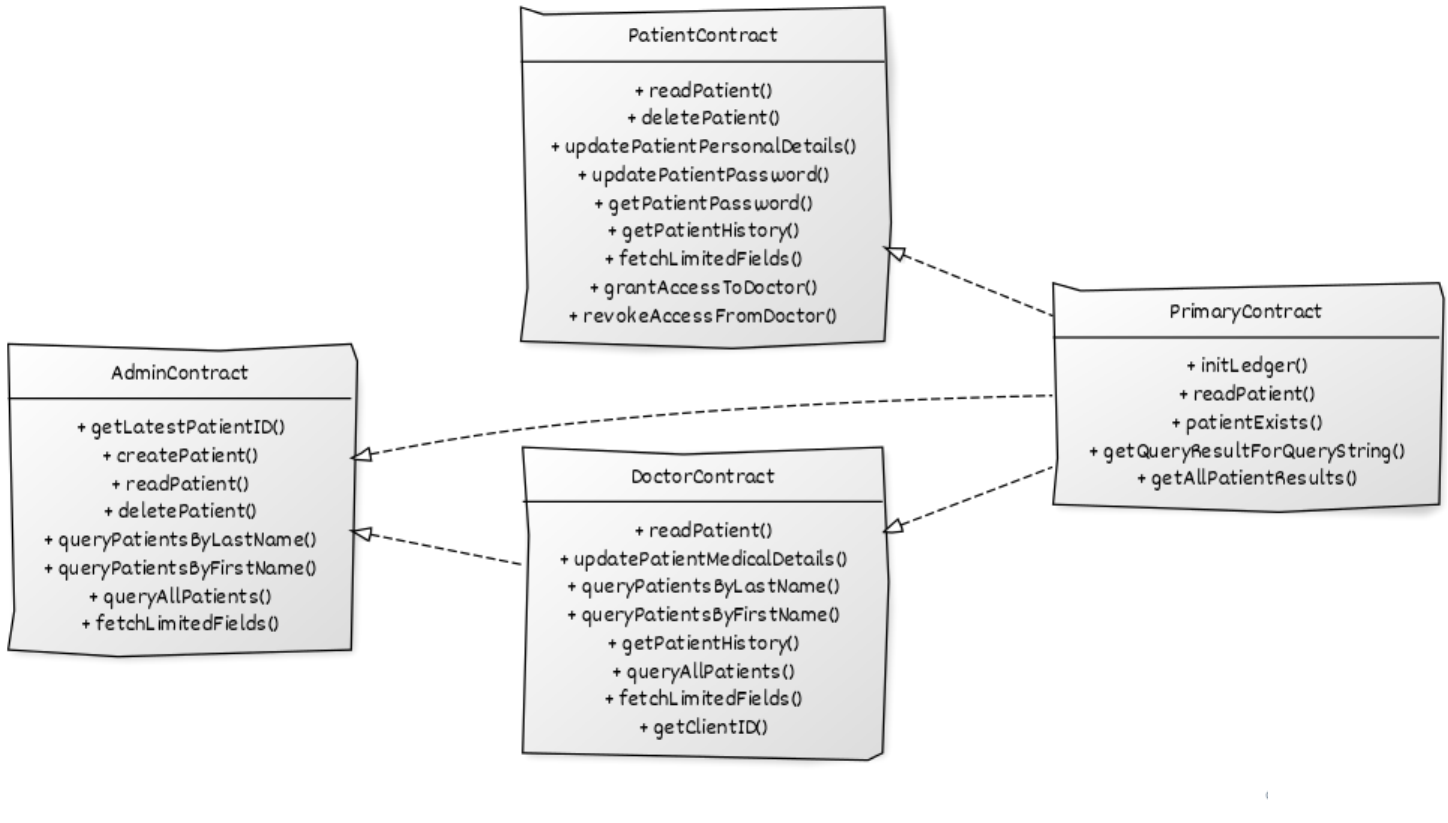}
	\caption{Relationships and Dependencies between the Contract Classes} \label{fig3}
\end{figure}

As depicted in Figure \ref{fig3}, there are three primary smart contracts packaged into a single \textit{chaincode}, each catering to specific roles (Admin, Patient, Doctor):
\begin{itemize}
	\item AdminContract: Invoked by the administrator, this contract endows the admin with the capability to create and delete patient records by adding or removing patient objects from the ledger. The admin can also retrieve information on all patients across the network.
	\item PatientContract: Designed for patient interactions with the ledger, this contract encapsulates logic tailored for the patient's role. Specifically, the patient can update and view personal details and passwords through the methods defined in the contract. Additionally, the patient contract includes methods to grant and revoke access from a doctor.
	\item DoctorContract: Tailored for doctor interactions, this contract provides methods enabling doctors to update and read patients' medical details.
\end{itemize}

The main reason for the three smart contracts is to ensure that each role has appropriate access to data on the ledger. Only the \textit{PatientContract} has the right to update personal details, grant or revoke access, and update the password. Similarly, the doctor's methods are inaccessible to the patient or admin. While the \textit{readPatient} method is common for all roles, the data retrieved by the contracts varies based on the role (i.e., the admin has access to patient names, the doctor has access to medical records, and the patient has access to the entire patient objects). This is to ensure a granular and role-specific access control mechanism within the system on the  Hyperledger Fabric. The breakdown of the key components of the hyperledger fabric in Algorithm \ref{alg:hyperledger} (Figure \ref{fig:hyperledger_algorithm})  illustrates the steps involve in setting up the network, deploying smart contracts, and performing transactions.

\begin{figure}[ht]
	\centering
	\begin{algorithm}[H]
		\caption{Setting Up Hyperledger Fabric Network and Smart Contracts}\label{alg:hyperledger}
		\begin{algorithmic}[1]
			\State \textbf{Initialization}
			\State Define Org1 (Hospital/Clinic) and Org2 (Patients)
			\State Create Peer0 for Org1 and Peer1 for Org2
			\State Establish Channel 1 for communication between peers
			\State Configure Membership Service Provider (MSP) for identity management
			\State Setup Certificate Authority (CA) for digital certificates issuance
			
			\State \textbf{Smart Contracts Development}
			\State Develop AdminContract, PatientContract, and DoctorContract using JavaScript
			\State Package contracts into chaincode
			
			\State \textbf{Network Deployment}
			\State Deploy chaincode onto the Hyperledger Fabric network
			\State Initialize CouchDB for world state storage
			
			\State \textbf{Transactions Management}
			\State \textit{AdminContract}
			\State Create and delete patient records
			\State Retrieve information on all patients
			
			\State \textit{PatientContract}
			\State Update and view personal details and passwords
			\State Grant and revoke access to doctors
			
			\State \textit{DoctorContract}
			\State Update and read patients' medical details
			
			\State \textbf{Access Control}
			\State Ensure role-specific access to data on the ledger
			\State Implement granular access control for \textit{Admin}, \textit{Patient}, and \textit{Doctor} roles
		\end{algorithmic}
	\end{algorithm}
	\caption{Algorithm for Setting Up Hyperledger Fabric Network and Smart Contracts}
	\label{fig:hyperledger_algorithm}
\end{figure}

\section{Discussion}
\begin{figure}[h]
	\centering
	\includegraphics[width=4in]{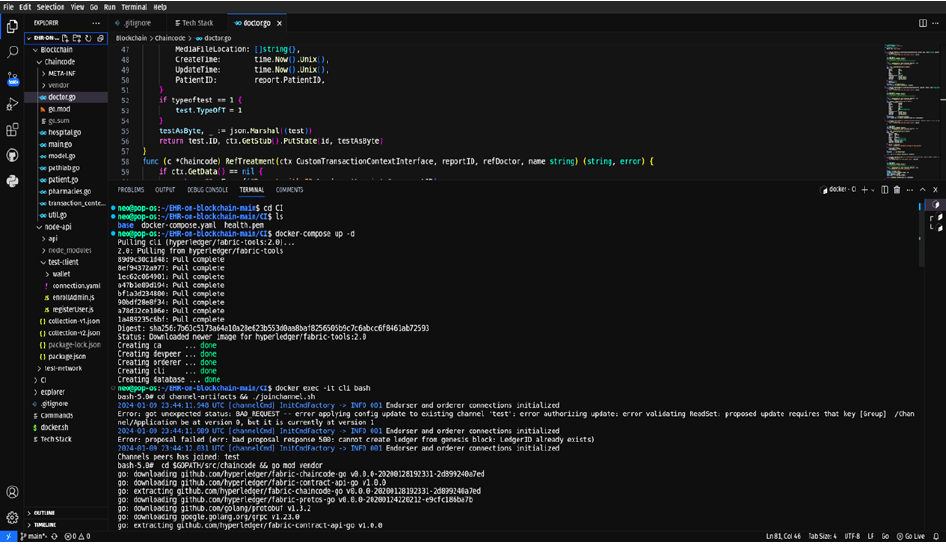}
	\caption{The Fabric Network Deployment} \label{fig5}
\end{figure}

The implementation of the blockchain-based Electronic Health Records (EHR) management system at Frere Provincial Hospital yielded promising results across various test cases involving three sets of users: administrators, doctors, and patients. Through rigorous testing, the system demonstrated its efficacy in enhancing data security, streamlining administrative processes, and improving patient care delivery. Figure \ref{fig5} shows a sample deployment of a local test fabric network, the installation and instantiation of a healthcare chaincode, and the visualisation of the transaction history using the Hyperledger Explorer.

\begin{figure}[ht]
	\centering
	\includegraphics[width=3in]{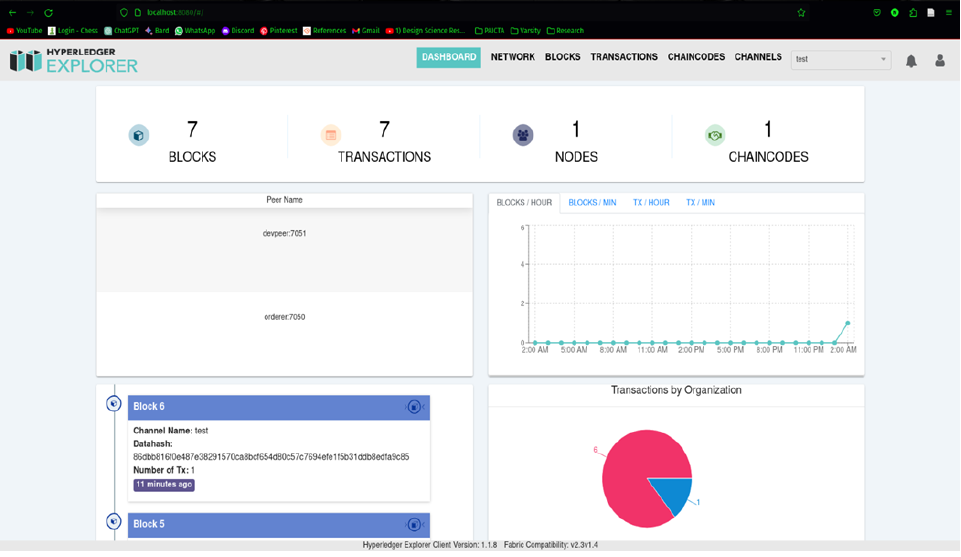}
	\caption{The Hyperledger Explorer Integration} \label{fig6}
\end{figure}

The testing was conducted in separate phases, specifically focusing on assessing the specific features designed for administrators, doctors, and patients. Every stage yielded valuable observations regarding the system's performance, usability, and impact on healthcare service delivery. Figure \ref{fig6} shows the complete Hyperledger Explorer integration.

Overall, our findings from comparing traditional record systems, analysing data from requirements gathering, and implementing the blockchain-based electronic health records (EHR) systems reveal the following:
\begin{itemize}
	\item Hyperledger Fabric's decentralised structure eliminates vulnerabilities associated with centralised traditional database systems.    
	\item The automatic correction mechanisms of Hyperledger Fabric enhance its effectiveness in mitigating manipulation risks. This provides reassurance regarding the integrity and authenticity of healthcare data, addressing concerns prevalent in traditional database systems.    
	\item Hyperledger Fabric's ledgers ensure data permanence, contrasting with the mutable nature of traditional databases. This highlights the significance of data integrity and permanence in healthcare settings, where accurate and reliable records are paramount.
\end{itemize}

\section{Conclusion}
The study examines the implementation of Hyperledger Fabric, a type of blockchain technology, for managing electronic health records at Frere Provincial Hospital. The study demonstrated the pragmatic application and advantages of blockchain technology in healthcare information systems.
The study offers insightful information about how blockchain technology can revolutionise healthcare information systems by enhancing EHR management efficiency, security, and privacy. Although our study highlights the potential use of blockchain technology in the management of electronic health records, it is important to acknowledge our research's limitations. As part of our ongoing exploration of this technology, the study is a preliminary investigation into the practical deployment and advantages of blockchain technology in healthcare settings. The findings, derived from a single case study, may not comprehensively encompass the intricacies and complexities of wider healthcare settings. In light of our practical implementation of the system, additional investigation and experimentation are necessary to thoroughly examine the complete functionalities and prospective uses of blockchain technology in healthcare settings. Specifically, challenges persist within the Hyperledger Fabric necessitating ongoing development, such as transaction history retrieval issues and performance concerns, which underscore the need for continued refinement. 

\subsection*{Acknowledgments}
\noindent The authors extend sincere gratitude to Frere Provincial Hospital for their collaboration and support during the implementation of the blockchain-based Electronic Health Records (EHR) system. The opinions expressed in this study are solely those of the research team.
%
%
%
%

\end{document}